\documentclass[runningheads]{llncs}
\usepackage[T1]{fontenc}
\usepackage{xspace}
\usepackage{cite}
\usepackage{amsmath,amssymb,amsfonts}
\usepackage{algorithmic}
\usepackage{graphicx}
\usepackage{textcomp}
\usepackage{xcolor}
\usepackage{bytefield}
\usepackage{kotex}
\usepackage{subcaption}
\usepackage{cleveref}
\usepackage{adjustbox}

\usepackage{pifont}
\crefname{figure}{Fig.}{Figs.}
\crefname{algocf}{Algorithm}{Algorithms}
\crefname{table}{Table}{Tables}
\crefformat{section}{#2\S#1#3}
\crefname{equation}{}{}
\begin{document}
\title{Field Testing and Detection of Camera Interference for Autonomous Driving}

\author{Kibeom Park\inst{1}\orcidID{0000-1111-2222-3333} \and
Huy Kang Kim\inst{2}\orcidID{1111-2222-3333-4444} }
\authorrunning{K. B. Park and H. K. Kim}
%
\author{Ki Beom Park \and Huy Kang Kim}
\institute{School of Cybersecurity, Korea University, Republic of Korea \\ 
\email{\{vkdnj0413,cenda\}@korea.ac.kr}}

\titlerunning{Field Testing and Detection of Camera Interference for Autonomous Driving}
%
\maketitle              
\begin{abstract}
In recent advancements in connected and autonomous vehicles (CAVs), automotive ethernet has emerged as a critical technology for in-vehicle networks (IVNs), superseding traditional protocols like the CAN due to its superior bandwidth and data transmission capabilities. This study explores the detection of camera interference attacks (CIA) within an automotive ethernet-driven environment using a novel GRU-based IDS. Leveraging a sliding-window data preprocessing technique, our IDS effectively analyzes packet length sequences to differentiate between normal and anomalous data transmissions. Experimental evaluations conducted on a commercial car equipped with H.264 encoding and fragmentation unit-A (FU-A) demonstrated high detection accuracy, achieving an AUC of 0.9982 and a true positive rate of 0.99 with a window size of 255. 

\end{abstract}

\section{Introduction}
Automotive ethernet~\cite{Automotiveethernet} in recent connected and autonomous vehicles (CAVs) stands for in-vehicle networks (IVNs) between electric control units (ECUs). controller area network (CAN) traditionally has dominated the IVNs market share.
However, due to the limited bandwidth of CAN, automotive ethernet is becoming a successor to enable high-definition applications such as video streaming for autonomous driving and infotainment systems. 

CAN supports a data transfer rate of up to 1 Mbps, but its limited bandwidth renders it incapable of handling simultaneous communications with multiple devices necessary for autonomous driving.

Conversely, automotive ethernet facilitates data transfer speeds of up to 1 Gbps, enabling real-time transmission of data from cameras, LiDAR and Radar. Its extensive bandwidth supports seamless communication with multiple devices. Furthermore, it utilizes protocols such as the Audio/Video Transport Protocol (AVTP), Scalable Service-Oriented Middleware over IP (SOME/IP) and Diagnostics over Internet Protocol (DoIP) to provide various functions essential for autonomous driving.

Autonomous driving technologies leverage an array of sensor equipment, including cameras, LiDAR and Radar, to detect pedestrians, interpret traffic signals, gauge distances from other vehicles, and recognize lane markings to maintain the lane. Recently, Tesla introduced a vehicle capable of autonomous driving using only cameras~\cite{TeslaVision}. This camera-based autonomous driving system, known as Tesla Vision, operates without ultrasonic sensors or LiDAR.
While automotive ethernet-based IVNs offer significant advantages from a cybersecurity perspective, some security risks and threats persist. For example, although a switched medium prevents the occupation of the entire bandwidth by a specific node, as seen in CAN buses, it still transmits packets from legitimate ECUs as well as potential attack nodes. 

\begin{figure}[h]
    \centering
    \includegraphics[scale=0.95,width=\textwidth]{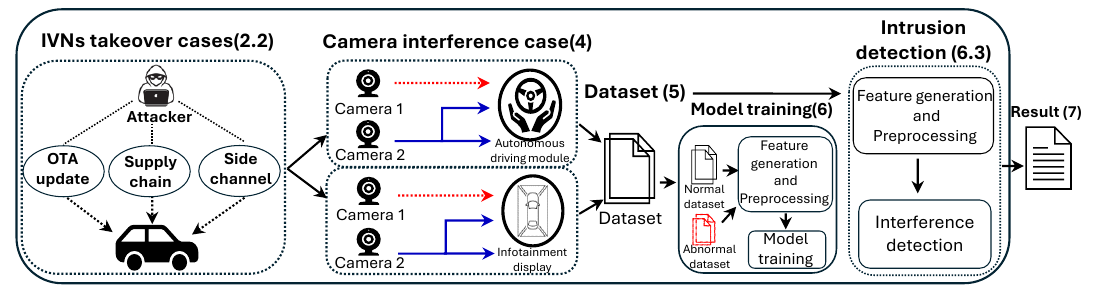}
    \caption{Overview of the proposed methodology for intrusion detection workflow.}
    \label{fig:total_pic}
\end{figure}

A compromised ECU can enable an attacker to manipulate the vehicle in desired ways, potentially causing significant harm to the driver and the surrounding environment. For instance, an attacker who has taken control of an ECU connected to the vehicle’s data transmission switch can disrupt the data being sent to the autonomous driving module. Among various attacks in automotive ethernet environment, we focused on camera interference attack (CIA) that can disrupt autonomous driving. The overall workflow for the CIA detection is as shown in~\cref{fig:total_pic}. We assumed that an attacker compromise the IVNs by exploiting vulnerabilities  of OTA, supply chain management or using physical access. Then, the attacker can send the CIA to the target car. Based on the attack scenarios, we built the datasets which includes the normal and attack cases. Then we developed a model for the CIA detection and estimated the overall performance of the model.

IDS is considered as a primary countermeasure to monitor in-vehicle traffic and detect anomalies. Several studies (see~\cref{subsec:related_work}) have recently proposed IDS to secure automotive ethernet-based IVNs. Although these proposed methods have exhibited satisfactory performance, a common limitation in the literature is that experimental results are based on datasets captured on a testbed rather than on commercial vehicles. 
~\vspace{-1cm}

\subsection{Contribution}
In this paper, we first provide a proof-of-concept of CIA on an automotive ethernet-driven vehicle, specifically, Hyundai Genesis G80 model, based on a malicious packet injection attack.
To the best of our knowledge, this is the first study on a commercial automotive ethernet-driven vehicle.
Second, we have compiled a network intrusion dataset containing benign traffic (from legitimate around-view cameras) and abnormal traffic (from the attacker).
Finally, we propose CIA-IDS, an IDS that effectively detects the CIA related hacking attempts. We leverage a novel feature to detect network intrusion on automotive ethernet.

\section{Background}
\subsection{Ethernet-Based IVNs}
Automotive ethernet has been introduced in modern vehicles to overcome the limitations in transmission speed and bandwidth of traditional IVNs such as CAN. By leveraging the high transmission speed and bandwidth of automotive ethernet, it is possible to transmit larger volumes of data in real-time for applications such as autonomous driving and infotainment systems.

Automotive ethernet includes SOME/IP, where one ECU requests events from other ECUs to exchange information; DoIP, which communicates with remote locations over the automotive ethernet to check the online status of an ECU; Additionally, AVTP is employed to handle time-sensitive data, ensuring synchronization of video playback through time synchronization mechanisms~\cite{IEEE1722}.
~\vspace{-0.5cm}

\subsection{Attack Vector}
\label{sec:Attack Vector}
We explain how the attacker accesses IVNs using our proposed method.
\begin{itemize} 
    \item \textbf{OTA Update Attack}
    OTA updates allow car manufacturers to update and enhance vehicular software remotely. exploiting OTA capability present a potential avenue for attackers to compromise vehicular control. Malefactors could upload a malicious update to the car manufacturer's OTA server, subsequently distributing this compromised update to the vehicles.
    \item \textbf{Supply Chain Attack}
    A supply chain attack involves adversaries infiltrating the automotive manufacturing process via specific components or software supply chains to embed malicious code or vulnerabilities. This subsequently allows for potential control over the car's internal network or control systems.
    \item \textbf{Physical Access}
    Physical approaches to gaining access method is also exist. Side-channel analysis is an attack technique used to extract information from a device that is not physically accessible. In cars that use smart keys, side-channel analysis allows attackers to detect and analyze the wireless communication signals of the smart key, enabling them to unlock the vehicle~\cite{side_attack}.
\end{itemize}

\section{Related Work}
\label{subsec:related_work}

Autonomous vehicles employ various methods to recognize the road conditions ahead and their surrounding environments, such as LiDAR sensors and V2X communication technology. However, cameras are also one of the frequently used technologies~\cite{camera_assum}. Consequently, recent studies have highlighted the potential risks of vehicle camera attacks and proposed detection methods.

Jeong~\textit{et al.}~\cite{jeong_AVTP} presents a method to detect replay attacks by manipulating the AVTP image data of vehicles in a testbed environment utilizing AVB Listener and AVB Talker. They released the `Automotive ethernet Intrusion Dataset'~\cite{jeong_data}. They utilized a CNN model as the detection approach for this dataset, achieving an accuracy of 0.9955. Their dataset contains payloads of the MPEG codec. 

Natasha~\textit{et al.}~\cite{alkhatib_AVTP} introduced a real-time IDS using deep learning and machine learning techniques, utilizing the automotive ethernet Intrusion Dataset~\cite{jeong_data}. They performed anomaly detection on the dataset based on an unsupervised deep learning model, the autoencoder. Utilizing these models for unsupervised learning on the~\cite{jeong_data}, they demonstrated performance metrics such as an F1-score of 0.98 for the LSTM autoencoder.

Han~\textit{et al.}~\cite{HAN_TOW} introduced a methodology for detecting abnormalities in automotive ethernet based on the Wavelet transform. This approach proves to be considerably more effective than methods utilizing ResNet or EfficientNet. However, it's worth nothing that the data used in this study was not extracted from vehicles in actual driving conditions.

Shibly~\textit{et al.}~\cite{shibly_AVTP} generate attack data using a generative adversarial network (GAN) based on the dataset from~\cite{jeong_data}. They use this dataset to conduct detection using a feature-aware semi-supervised approach.

Jeong~\textit{et al.}~\cite{jeong_AERO} proposes an IDS named automotive ethernet Real-Time Observer (AERO), which analyzes traffic from various protocols in IVNs. This is an unsupervised IDS that learns only from normal condition data and detects anomalies.

Song~\textit{et al.}~\cite{Review_attack_defence} explain the network attacks and defense methods that can occur in CAN networks and automotive ethernet.

Choi~\textit{et al.}~\cite{driving_pattern_rnn} and Han~\textit{et al.}~\cite{driving_pattern} propose models that can identify drivers based on patterns such as acceleration and braking, which are indicative of driving habits, using deep learning models.

\section{Proof of Concept}
\begin{figure}[h]
    \centering
    \includegraphics[scale=1]{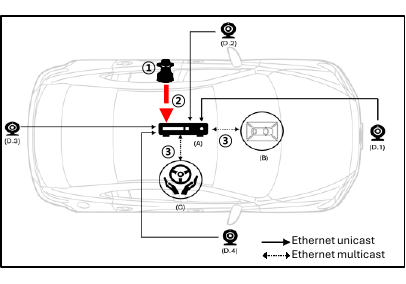}
    \caption{Example for attack autonomous driving module. (A) is ethernet switch, (B) is infotainment display, (C) is autonomous driving module, D.1, D.2, D.3, D.4 are on-board camera.}
    \label{fig:attack_arch}
\end{figure}

An attacker can exploit previously introduced methods (\cref{sec:Attack Vector}) to compromise an ECU connected to an ethernet switch responsible for data transmission within IVNs utilizing automotive ethernet.

By taking control of a compromised ECU, the attacker can manipulate data transmitted to the autonomous driving module.~\cref{fig:attack_arch} illustrates this process.
Step ① represents the attacker's intrusion into the IVNs, step ② indicates the attacker's control over the ECU connected to the ethernet switch within the IVNs, and finally, step ③ shows the attacker executing an attack on the infotainment display and autonomous driving module.

The ethernet switch is responsible for transmitting data sent by ECUs or on-board cameras to their appropriate destinations. When necessary, the switch uses multicast transmission to efficiently send data to multiple recipients within a specific group. This is particularly effective for modules such as the autonomous driving module or infotainment system, which require large amounts of data.

In this network environment, we execute an attack to disrupt the transmission of camera image data to the autonomous driving module or the infotainment system.
Specifically, we targeted the infotainment display to demonstrate an attack where camera image data is transmitted abnormally. This abnormal data transmission attack can be explained by a cache poisoning attack~\cite{cache-poisoning}. Similarly, the autonomous driving module receives image data using multicast transmission, just like the infotainment display. Therefore, by employing the same attack method on the autonomous driving module, we can disrupt the normal transmission of camera image data to the module.
\begin{figure*}[t]
    \begin{subfigure}[t]{0.32\textwidth}
        \centering
        \includegraphics[scale=0.85]{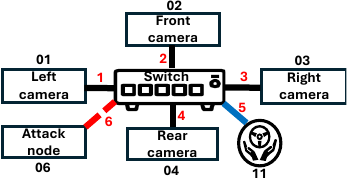}
        \caption{Example of ethernet switch.}
        \label{fig:poc_switch}
    \end{subfigure}
        \hfill
        \centering
    \begin{subfigure}[t]{0.25\textwidth}
        \includegraphics[scale=0.83]{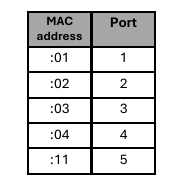}
        \caption{MAC table before cache poisoning attack.}
        \label{fig:table1}    
    \end{subfigure}
        \hfill
        \centering
    \begin{subfigure}[t]{0.25\textwidth}
        \includegraphics[scale=0.83]{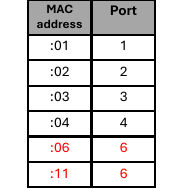}
        \caption{MAC table after cache poisoning attack.}
        \label{fig:table2}    
    \end{subfigure}
    \caption{Cache poisoning attack in vehicle.}
    \label{fig:cpa_pic}
\end{figure*}

\begin{figure}[t]
    \centering
    \begin{subfigure}{0.49\textwidth}
        \includegraphics[scale=0.2]{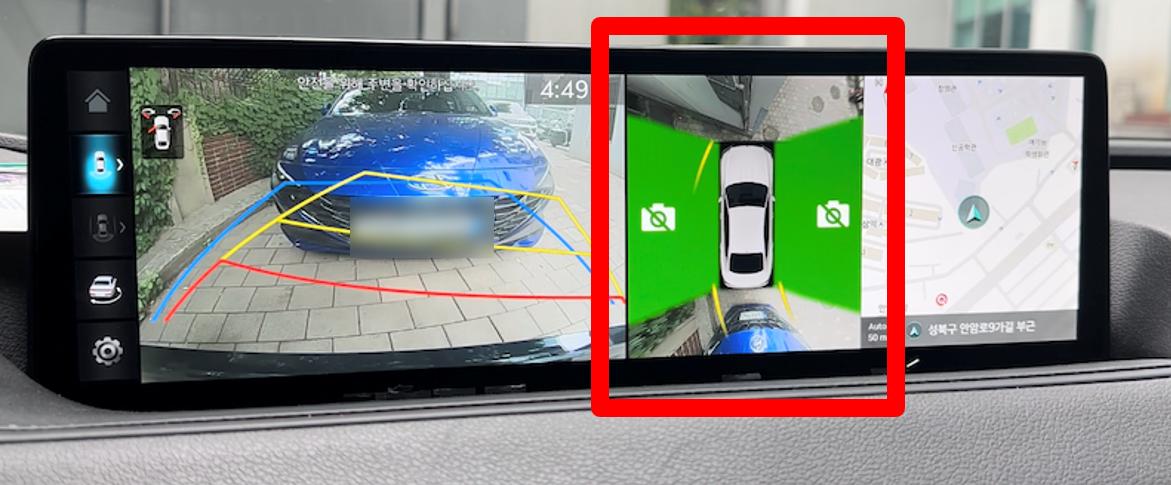}
        \caption{Under normal status}
        \label{fig:before_attack}
    \end{subfigure}
    \hfill
    \centering
    \begin{subfigure}{0.49\textwidth}
        \includegraphics[scale=0.2]{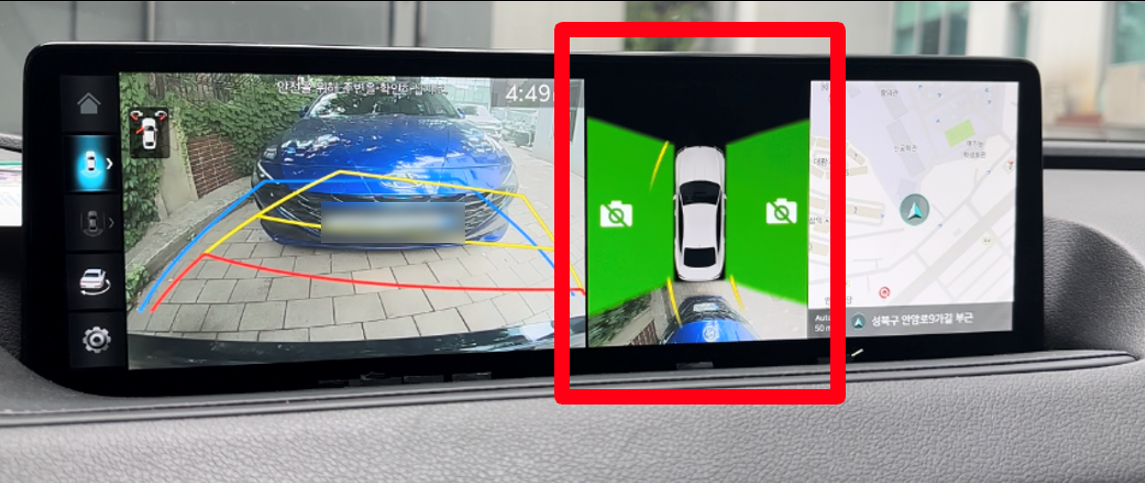}
        \caption{Under CIA status}
        \label{fig:after_attack}
    \end{subfigure}
    \caption{Demonstration of the camera inference attack. The infotainment system becomes blind when it goes under the attack.}
    \label{fig:attack_result}
\end{figure}

\cref{fig:cpa_pic} illustrates the switch architecture and changes in the MAC address table that enable abnormal data transmission to the autonomous driving module using a cache poisoning attack. 
An attacker who has compromised an ECU connected to the ethernet switch manipulates data destined for the MAC address at port 11, redirecting it through port 5 to themselves. After intercepting the data, the attacker mixes it with other data before transmitting it to the autonomous driving module.
As shown in the "\textbf{Camera interference case}" part of ~\cref{fig:total_pic}, we transmit the data from one camera to an incorrect destination. 
To carry out this attack, we use the Rad-Galaxy equipment, which is capable of sniffing the vehicle's data. This equipment allows us to intercept and analyze the data traffic within the vehicle network, facilitating the execution of the attack.
The Rad-Galaxy is a multipurpose ethernet tap and media converter tailored for automotive ethernet applications~\cite{rad-galaxy}. 
We configure the direction of the left camera data towards the front display using the Gateway Builder~\footnote{Rad-Galaxy program that changes the data transmission path of IVNs.} tool provided by this device.
\cref{fig:attack_result} shows the results of executing a CIA on an actual vehicle.
~\cref{fig:attack_result}(a) depicts the normal operation of the cameras, with both the front and left cameras working correctly. 
~\cref{fig:attack_result}(b) shows the outcome after attempting an attack using the Rad-Galaxy equipment.
\section{Dataset}
\subsection{Dataset Description}
\label{sec:Dataset Description}
In this section, we describe the differences between the normal and anomalous datasets. We extract the data by using the rad-galaxy tool. The normal dataset comprises packets transmitting front camera data to the front display. The anomalous dataset includes packets where data from both the left-side and front cameras are simultaneously transmitted to the front display. 

The normal dataset, collected over 30 minutes of driving, is approximately 5.86GB, containing 6,346,876 packets. The anomalous dataset, also gathered during a 30-minute drive on campus, is approximately 4.97GB, containing 5,450,620 packets. The images captured by the front, left, right and rear cameras are transmitted to their respective displays. Normal dataset consisting solely of front-camera image data are labeled as 0, while attack dataset containing both front and left-side image data are labeled as 1. As the goal is to detect deviations from this consistency, labeling all attack dataset, which include both frontal and left-side mixed attacks, as 1 can yield favorable results.

The image data captured by the camera is too large to be transmitted in a single packet, so it is fragmented before transmission. In our experiment, we use the FU-A fragmentation method~\cite{FU_A}. 
Fragmentation ensures that packets do not exceed the MTU size. In a normal dataset, the fragmented packets create continuity. However, due to our attack, extraneous data are injected into the normal dataset, disrupting this continuity. As a result, the continuity observed in a normal dataset is altered.
To demonstrate the difference in length sequence between normal and abnormal datasets, we utilize a heatmap for visualization.

\begin{figure*}[t]
    \begin{subfigure}[t]{0.45\textwidth}
        \centering
        \includegraphics[scale=0.5]{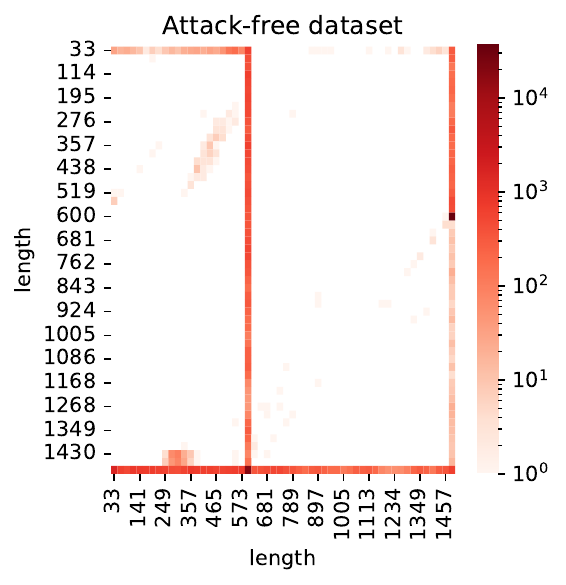}
        \caption{Heatmap for normal dataset length sequence.}
        \label{fig:length_attack_free}    
    \end{subfigure}
        \hfill
        \centering
    \begin{subfigure}[t]{0.45\textwidth}
        \includegraphics[scale=0.5]{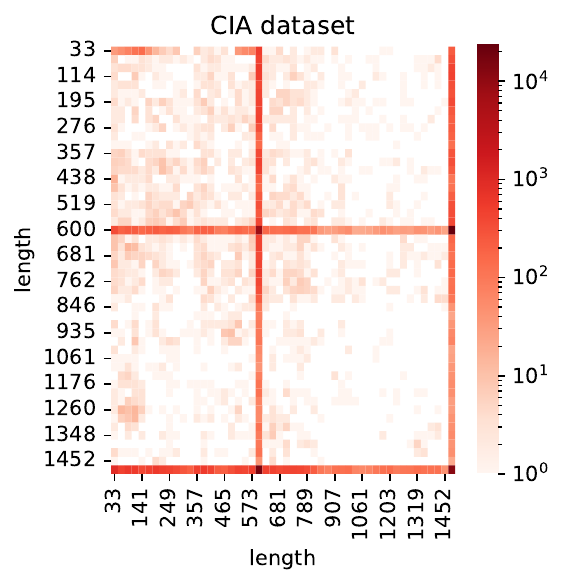}
        \caption{Heatmap for abnormal dataset length sequence.}
        \label{fig:length_attack}    
    \end{subfigure}
    \caption{Heatmap for length sequence 2-gram feature.}
    \label{fig:length seqeunce}
\end{figure*}

~\cref{fig:length seqeunce} illustrates the differences in the length sequences between the normal and abnormal dataset. Each figure represents the 2-gram of the length sequences from the datasets. For example, for the 576, 1474, 160, 1474 sequence, 2-grams of $P = \{[576, 1474], [1474, 160], [160, 1474]\}$ can be expressed, and $P_i[0]$ is the X-axis. We can determine the frequency by specifying $P_i[1]$ as the Y-axis. Darker colors indicate a higher frequency of the corresponding 2-gram. It is evident that there are differences in the 2-gram patterns between the normal and abnormal datasets.

We employ data length as a feature that represents these characteristics and describe the method used for anomaly detection utilizing this feature.
To effectively train the model on the continuity of packets, we need to determine a window size. 
The packet's sequence number has a maximum value of 255. Under normal status, the packet sequence increases sequentially from 0 to 255. 
We chose a window size corresponding to the packet's sequence number, ranging from 0 to 255.
When an attack occurs, this sequence within the range of 0 to 255 displays irregular continuity. 
\section{Methodology for CIA Detection}
\label{sec:methodology}
We described potential attacks on vehicle cameras using automotive ethernet and the differences between datasets in normal and attack scenarios.
In this section, we introduce the detection system for CIA.
We propose a methodology for the model to learn the sequence of length values described earlier and detect sequences that differ from the learned ones.

\subsection{Gated Recurrent Unit (GRU)}
Utilizing the time-series data preprocessing, we employ a GRU deep learning model to detect camera interference attack in vehicle cameras \cite{GRU}.

The GRU model is specialized for processing time-series and sequential data. It operates using two key mechanisms. The update gate is used to determine how much past information to retain, whereas the reset gate decides how much past information to forget. Leveraging these characteristics, the model learns information about the sequence of payload data lengths, determining at which point to start forgetting, and conducts anomaly detection based on the learned content.
The following equations govern the behavior of a GRU:
\begin{equation}
    r_{t}=\sigma (W_{r}\cdot [h_{t-1},x_{t}] + br)
\end{equation}
\begin{equation}
  z_{t} = \sigma (W_{z}\cdot [h_{t-1},x_{t}]+b_{z})  
\end{equation}
\begin{equation}
  \widetilde{h_{t}} = tanh(W\cdot [r_{t} * h_{t-1},x_{t}] + b)  
\end{equation}
\begin{equation}
  h_{t} = z_{t}*h_{t-1} + (1-z_{t})*\widetilde{h_{t}}  
\end{equation}
The reset gate's weight $W_{r}$ is operated with the previous state information $h_{t-1}$ and the current state information $x_{t}$ to determine how important $h_{t-1}$ and $x_{t}$ are. This operation involves an activation function. $r_{t}$ represents the reset gate, which determines how much of the past information to forget in the computation process. This formulation allows the determination of how much of the past information to forget. The update gate operates on the previous state information and the current state information using the update gate's weights, followed by the application of an activation function. Using the reset and update gates, it is decided how much of the previous information to retain or forget.
To achieve this, weights are computed for the current state, and a hyperbolic tangent activation function is used to express the weights within the range of -1 to 1. Finally, the update gate $z_{t}$ is employed to determine how much information to blend between the previous state information $h_{t-1}$ and the current state information ${h_{t}}$.
\subsection{Data Preprocessing} 
\begin{figure}[h]
    \centering
    \includegraphics[scale=0.38]{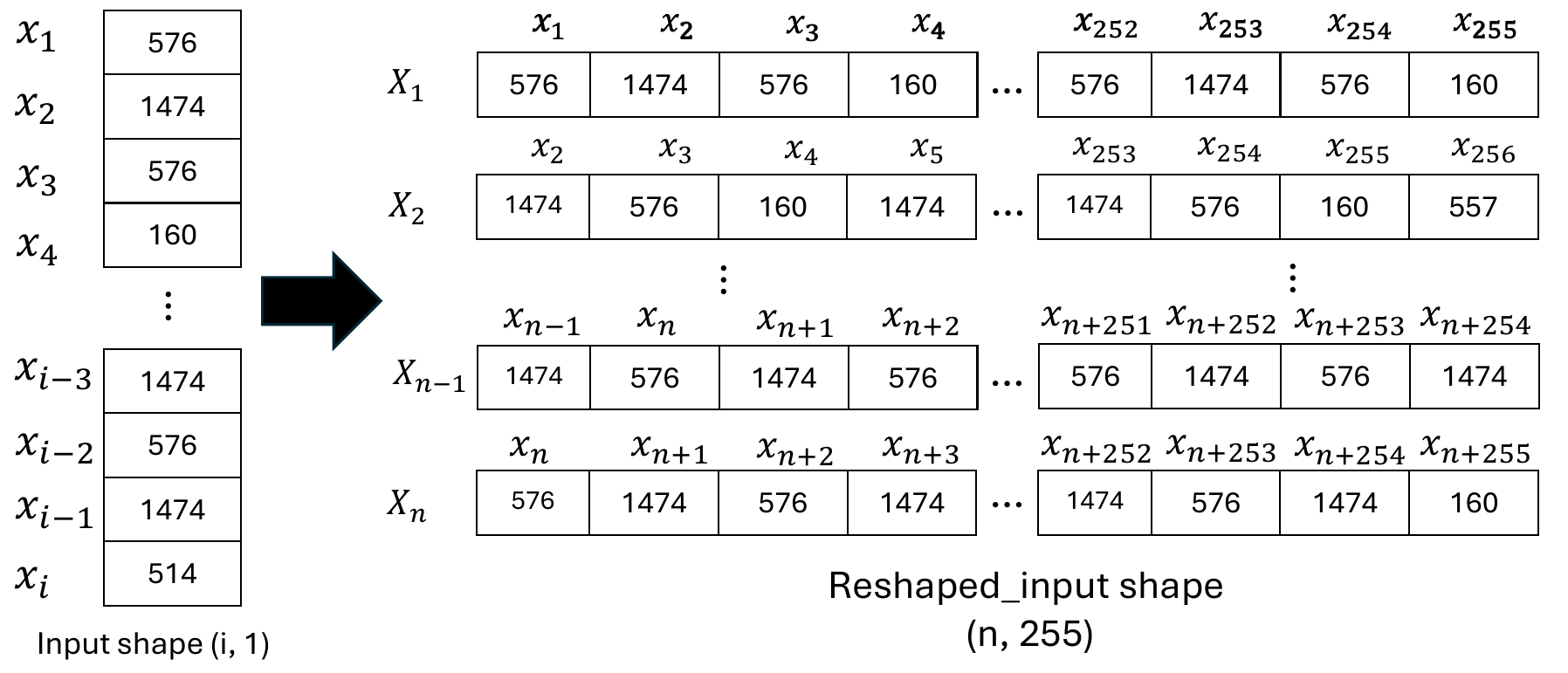}
    \caption{Dataset reshaping for model input.}
    \label{fig:reshape}
\end{figure}
For the model to be properly trained and accurately distinguish anomalies, preprocessing of the dataset is necessary.
In section IV, we describe the dataset we extracted for our experiments. 
This dataset comprises camera data continuously extracted over time. 
Such data is referred to as time-series data, and while a single data point can be meaningful, analyzing a sequence of data can yield more significant information.
As mentioned in~\cref{sec:Dataset Description}, our dataset's data length encompasses the time-series information arising from fragmenting continuously captured image data. 

In this section, we describe the preprocessing process for the dataset to ensure that the model accurately learns this consistency.
We denote each packet's data length as $\textit{x}_{i}$. A single window, segmented by the window size, is represented as $\textit{X}_{n}$.
Since a window comprises 255 rows, a single window can be expressed as $X_{n} = {x_{n} ... x_{n+254}}$. 
However, training the model in this form would only teach it to learn from a single payload data length, not a sequence.
To train the model on a window containing 255 packets, we reshape the 255 rows into a single row composed of 255 columns.
\cref{fig:reshape} illustrates the process of reshaping through a diagram. First, input shape for i in (i, 1) means the number of total packets. 
We reshape this input shape to (n, 255). Here, n represents the number of windows.

Preprocessing is also required for the label indicating whether the dataset is normal or anomalous. 
Like the data length, labels are also divided according to a window size of 255. 
We can represent the label for an individual packet as $y_{1}$. A set of labels divided by the window size, $Y_{i}$, has 255 labels and is composed as $Y_{i}={y_{1} ... y_{255}}$. To represent $Y_{i}$ as a single label for the 255 labels, we use the average value of these 255 labels.

\subsection{GRU-Based Camera Interference Attack IDS}
We describe a model capable of detecting attack using the previously introduced feature selection, data preprocessing and GRU model. 

\textbf{Input Layer:} As explained in \cref{fig:reshape}, the shape of the dataset for input is (255, 1). Each $X_{i}$ contains values from $x_{i}$ to $x_{i+255}$, resulting in 255 columns. This input shape allows the GRU model to analyze the continuity of the 255 values of $x_{i}$.

\textbf{GRU:} GRU learns the continuity of the input $X_{i}$. The update gate determines how much information about the continuity of the current input data length should be retained. In contrast, the reset gate decides how much of the previously learned information about data length continuity should be forgotten.

\textbf{Output:} Based on the GRU model's learning results, the output layer produces predictions about the input. By comparing the predicted results with $Y_{i}$, the model can determine whether it has correctly detected normal sequences as normal and abnormal data as anomalous.
\section{Experiment Result}
\label{sec:Experiment}
\label{sec:experimental_results}
~\vspace{-0.9cm}
\begin{figure*}[ht]
    \begin{subfigure}[t]{0.3\textwidth}
        \centering
        \includegraphics[scale=0.45, height=2.5cm]{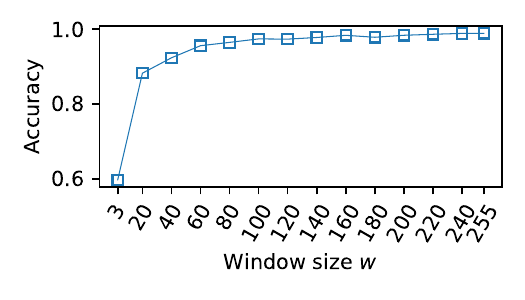}
        \caption{Detection accuracy}
        \label{fig:acc-graph}
    \end{subfigure}
        \hfill
        \centering
    \begin{subfigure}[t]{0.3\textwidth}
        \includegraphics[scale=0.45, height=2.5cm]{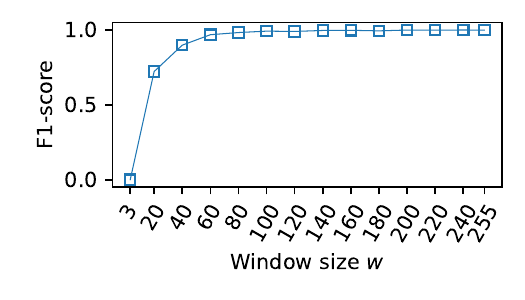}
        \caption{F1-score}
        \label{fig:f1-graph}    
    \end{subfigure}
        \hfill
        \centering
    \begin{subfigure}[t]{0.3\textwidth}
        \includegraphics[scale=0.45, height=2.5cm]{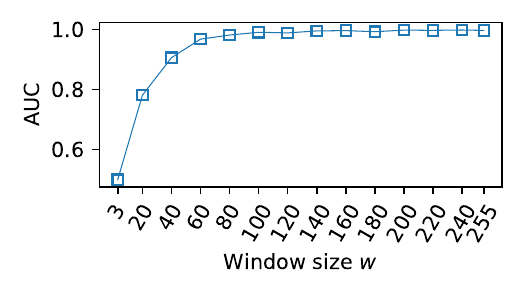}
        \caption{AUC}
        \label{fig:auc-graph}    
    \end{subfigure}
    \caption{Performance of CIA-IDS by window size}
    \label{fig:dif_window_graph}
\end{figure*}

\vspace{-0.9cm}
\begin{figure*}[h]
    \begin{subfigure}[t]{0.45\textwidth}
        \centering
    \includegraphics[scale=0.45]{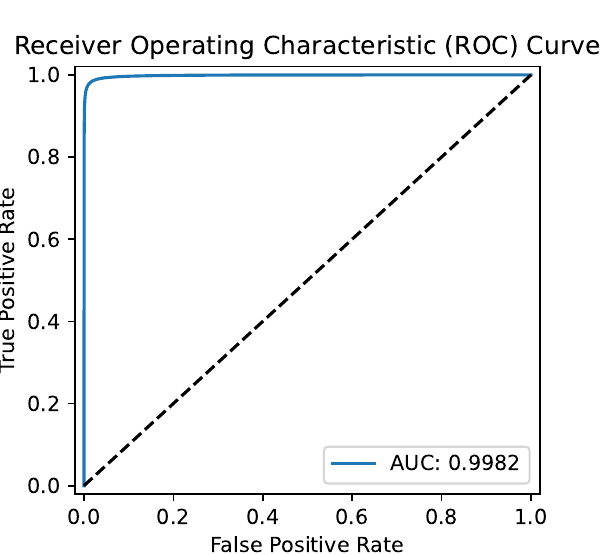}
    \caption{ROC curve}
    \label{fig:ROC_CURVE}
    \end{subfigure}
        \hfill
        \centering
    \begin{subfigure}[t]{0.45\textwidth}
        \includegraphics[scale=0.5]{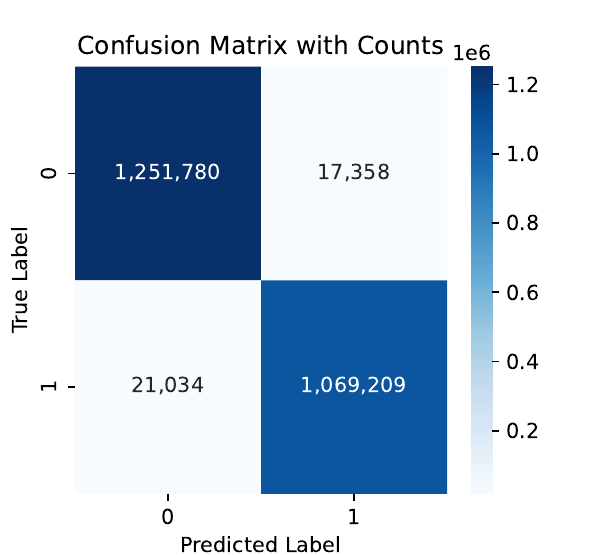}
        \caption{Confusion matrix}
        \label{fig:matrix}    
    \end{subfigure}
    \caption{ROC curve and confusion matrix}
    \label{fig:roc_conf}
\end{figure*}
In this study, we elucidate the detection results of abnormal camera data transmission scenarios using a GRU-based IDS facilitated by the previously described sliding-window data preprocessing technique. Our experimental setup encompasses an Intel i7-10700k CPU@3.80GHz, 64GB of memory and an NVIDIA RTX 3070 GPU, which are utilized for both training and evaluation. For the dataset generated through our preprocessing steps, we allocate 80\% of the data for training and 20\% for testing purposes. We explore various window sizes to assess detection performance alongside the impact of window size on model complexity. \cref{fig:dif_window_graph} illustrates the model's accuracy, F1-score and AUC for window sizes ranging from 3 to 255, in increments of 20.

In our experiments, we observe that despite the satisfactory performance at a smaller window size, a comparison of the inference times at window sizes of 140 and 255, yielding 0.002ms/$X$ and 0.009ms/$X$ respectively, shows negligible difference. Therefore, we have selected a window size of 255 for our IDS. ~\cref{fig:roc_conf}(a) presents the ROC curve for a window size of 255, exhibiting an AUC value of 0.9982.

Furthermore,~\cref{fig:roc_conf}(b) demonstrates the detection outcomes of the IDS on the test dataset, where the X-axis represents the model's predicted results, whether the window is classified as normal or containing an attack, and the Y-axis indicates the actual labels of the windows, whether they are truly normal or anomalous, forming a confusion matrix. The true positive rate, which achieves a value of 0.99, indicates outstanding performance.~\cref{table:performance_metrics} shows the performance metrics of the IDS.
~\vspace{-0.3cm}
\begin{table}[h]
\caption{Performance Metrics of the CIA-IDS}
\label{table:performance_metrics}
\centering
\begin{tabular}{|c|c|c|c|c|c|c|}
\hline
Accuracy & AUC     & F1-score & FPR    & FNR    & TPR    & TNR    \\ \hline
0.98848  & 0.99823 & 0.99824  & 0.0137 & 0.0193 & 0.9807 & 0.9863 \\ \hline
\end{tabular}
\end{table}
~\vspace{-0.9cm}

\section{Conclusion}
\label{sec:conclusion}
In this study, we successfully demonstrated a proof-of-concept for detecting CIA on a Hyundai Genesis G80, leveraging a GRU-based IDS designed for automotive ethernet.
Our approach utilized a novel sliding-window data based on packets`s length value sequence. The experimental results showed that our IDS, with a window size of 255, achieved high detection accuracy, an impressive AUC of 0.9982 and a true positive rate of 0.99, indicating its exceptional ability to differentiate between normal and anomalous data.
 It has the advantage of carrying out attacks targeting actual vehicles and using the corresponding data, but has the disadvantage of being able to apply IDS only in a network structure fragmented by the FU-A method and using H.264 encoding. In upcoming works, limitations can be overcome if the actual image is recovered using the payload of the data set and detection is performed based on the image.
~\vspace{-0.3cm}
\subsubsection{Acknowledgement.}
This work was supported by the 2021 Autonomous Driving Development Innovation Project of the Ministry of Science and ICT, Development of technology for security and ultrahigh-speed integrity of the next-generation internal network of autonomous vehicles under Grant 2021-0-01348.


~\vspace{-0.8cm}
\bibliographystyle{splncs04}
\bibliography{main}
\end{document}